\documentclass[twocolumn,aps,showpacs]{revtex4}

\begin{document}
\title{Atomic Force Microscopy of height fluctuations of fibroblast cells
}
\author{B\'alint Szab\'o, D\'avid Selmeczi}
\affiliation{Department of Biological Physics, E\"otv\"os University, Budapest
}
\author{Zsuzsanna K\"ornyei, Em\'{\i}lia Madar\'asz}
\affiliation{Institute of Experimental Medicine, Budapest, Szigony u. 43, Hungary
}
\author{No\'emi Rozlosnik}
\affiliation{Department of Biological Physics, E\"otv\"os University, Budapest, P\'azm\'any P\'eter s\'et\'any 1/A, 1117 Hungary
}
\date{\today}

\begin{abstract}
We investigated the nanometer scale height fluctuations of 3T3 fibroblast cells with the atomic 
force microscope (AFM) under physiological
conditions. Correlation between these fluctuations and lateral cellular 
motility can be observed. Fluctuations measured on leading edges 
appear to be predominantly related to actin polymerization-depolymerization 
processes. We found fast (5 Hz) pulsatory behavior with 1--2 nm amplitude on 
a
 cell with low motility showing emphasized structure of stress fibres. 
Myosin driven 
contractions of stress fibres are thought to induce this pulsation.
\end{abstract}


\maketitle

\section{Introduction
}
The motility of animal cells is dominated by actin-myosin-based
contraction and actin polymerization-based protrusion. The two basic types of protrusions, lamellipodia and filopodia
are driven by actin polymerization-depolymerization
processes \cite{a1,a2,a3}. The physical theory of such biological
motilities employing the tools of statistical physics \cite{a4,a5} gives
an idea how it works but it still needs to be developed.

A number of cellular activities can cause height fluctuations on the time and distance 
scales that we investigated. Actin polymerization and actin-myosin based contractions 
represent only one class. The rearrangement of the structure built up from 
intermediate filaments (IF) under the plasma membrane can be another source 
of vertical fluctuations. IF-s provide mechanical stability to animal cells. Any 
significant weakening of the IF array alters at least locally the elastic properties 
of the cell leading to increased susceptibility to intrinsic or extrinsic forces. 
Assembly or disassembly of large protein complexes in the plasma membrane or their 
lateral motion under the AFM tip will also result in vertical motility, not to 
mention endo- and exocytotic activity. Intracellular transport processes can have 
an effect on the vertical fluctuations, as well.

Cellular motility on the micrometer scale has been extensively
investigated with video-microscopy (\cite{a6,a7,a8} and references therein). Spatial
resolution provided by optical microscopy of living cells, however does not
enable researchers to observe nanometer scale motion and rearrangement
of cell components. AFM is an adequate tool for such measurements
\cite{a9,a10,a11,a12}. Stress fibres (contractile bundles of actin filaments and myosin-II)
play an important role in the control of cell shape and the adhesion of 
cells to the extracellular matrix
 through focal contacts. These
characteristic cytoskeletal elements can be imaged with the AFM due to their high
elastic modulus \cite{a13,a14}.
 AFM is
 capable not only for recording 
high-resolution topographic images of living cells
 but also for measuring the 
elastic properties of them simultaneously \cite{a15} and investigating cellular 
dynamics \cite{a16,a17}.

By repetitively scanning on the surface of a cell time-lapse images can
be recorded \cite{a18,a19}. The analysis of subsequent images yielding a movie
is highly informative in terms of the kinetics of the cytoskeleton. Although 
fast cellular motility can not be
 examined by the repetitive scanning procedure 
due to the minute-range of scanning time, nanometer
 scale fast motion can be 
probed by positioning the tip on
 the area of interest. In this way vertical 
fluctuations can be
 investigated almost without a limitation of the time scale. 

Close to the molecular scale the motion of cells is dominated by stochastic 
fluctuations. With the AFM we tried to shed light on the linkage between 
microscopic fluctuations and organized motility.

\section{Materials and methods}

\subsection{Cell Culture}

3T3 mouse fibroblasts were cultured in DMEM supplemented with 10\% fetal
calf serum (GIBCO), 100 units/ml penicillin G, 0.1 mg/ml streptomycin
and 0.75 $\mu$g/ml Amphotericin B (SIGMA) at 37 $^{\circ}$C, in
a 5\% CO$_2$ atmosphere. A few days before AFM measurements cells were
subcultured on 13-mm glass cover slips. 

\subsection{Atomic Force Microscopy
}

We investigated the nanometer scale motion of 3T3 mouse fibroblast
cells in culture. Different cells (C6 rat glioma) giving similar
--- unpublished --- results  were also examined. 

A commercial AFM (TopoMetrix Explorer, Santa Clara, CA) with custom-made
sample heating control system and fluid chamber was used. Measurements
were carried out at 37 $^{\circ}$C in CO$_2$ independent medium
containing 10\% fetal calf serum (GIBCO). We used soft silicon nitride
cantilevers (Thermomicroscopes, Coated Sharp Microlevers, Model \#
MSCT-AUHW, with typical force constant 0.01--0.03 N/m, 20 nm radius of
curvature). Topographic and deflection images were acquired in contact
mode. High-resolution images were acquired at a scanning frequency of
$\sim$4 Hz.  Non-destructive low force scanning provided stable
sustained imaging of living cells for 8--10 hours. After AFM experiments
cells were maintained in the same medium for 1--2 days and found to be normal.
We could not achieve high-resolution imaging on a portion of cells due to 
their increased height or softness. 

Considering that small details of cellular components can be observed
at the best quality on shaded deflection mode images we present our
experimental data in this format. Topographic images provide height
information but with poor contrast.

On the basis of consecutive images local lateral velocity of cells was calculated. At the
edges of cells we measured the average displacement of the contour. In
the middle of cells we chose some structures with characteristic shape
and the lateral displacements of these structures were measured.   

The measurement of local height fluctuations of cells was started at least 1/2
hour after mounting the sample into the fluid cell. In this way thermal
transient effects could be eliminated. After each scanning the tip was
positioned onto the point of interest with the same force and feedback
parameters  and we captured the DC voltage of the z-piezo by a digital
oscilloscope  (Tektronix TDS 210) for 22.5 seconds with 100 Hz sampling rate. 

\section{Results}

Distinct types of cellular motility could be examined by the measurement of 
vertical fluctuations.

FIG.\ \ref{fig1} displays 2 images of a movie showing the slight motility
of the rear edge of a cell with a 4.5 minutes time shift. This cell was almost
quiescent during the experiment with a highly stable structure of cytoskeletal
fibres and moderate lateral motility. The rear edge is being pulled by the 
stress fibres: see the parallel set of curved fibres anchored to the edge of 
the cell. In the same time cell-matrix junctions or nonspecific contacts 
adhering the rear of the cell to the support weaken and break. We also observed 
a typical retracting triangular shaped ~20 $\mu$m wide contact (image not shown) 
of the same cell at the rear edge. The contact was broken a few minutes after 
recording the vertical fluctuations.

Typical vertical fluctuations registered on these two locations are
presented in FIG.\ \ref{fig2}. We suppose that the apparent difference
between vertical fluctuations originates in the different
biological activities of the two regions. While the entire region of the cell
shown in
 FIG.\ \ref{fig1} was extremely stable with a lateral velocity of about 2
nm/s, the edge beside the retracting triangular shaped contact moved
with a speed of about 11 nm/s.

To analyze height fluctuations we calculated the power spectrum and the
height-height correlation function with a maximal $\tau$=5 s time shift of each $x(t)$ height-time curve: 

\begin{equation}
y^2(\tau)=\sum_{t}\frac{\left( x(t)-x(t+\tau)\right) ^2}{N},\ t=i\Delta t,\ i=1..N,
\end{equation}

\begin{equation}
N\Delta t=22.5-5 s
\end{equation}

\noindent 
where $\Delta t$ (10 ms) is the sampling time.

This function can characterize stochastic height fluctuations by giving the average change of height as a function of time.  Curves are presented in FIG.\ \ref{fig3}, the number of measured 
height-time
 curves $n$ is indicated. The lateral velocity of each location seems to 
correlate with the saturation value of the height-height correlation function measured
at that location confirming our assumption that height fluctuations are related to 
local
 biological activity (motility). The starting slopes of the curves give
the speed of fast fluctuations. Curves saturate with different
characteristic (saturation) times. There is an apparent difference between
curves (a) and (b) in the saturation value. The
 characteristic time ($\sim$2 s) of 
curve (a') is approximately double those of the
 other two curves. Curve (a') 
was registered on the middle region (cell body) of the quiescent cell. (See 
Table \ref{table1}.) Characteristic time and
 saturation value are related to the 
average duration and amplitude
 respectively of an upward or downward motion.

The analysis of power spectra (FIG.\ \ref{fig4}) of the height-time curves acquired 
on each location of this quiescent cell revealed sustained periodic fluctuations 
during the experiment (1.5 hours). We found a 
characteristic peak at 4.9 Hz with a 
width of 3.5 Hz. The area of this peak gives an average amplitude of 1.5 $\pm$ 0.4 nm. 
Cells without apparent stress fibres nearby 
lack this peak.  The origin of the sharp 
peaks in the spectrum is 
electric noise. 

FIG.\ \ref{fig5} shows the contours of a leading edge of a motile cell from consecutive images. Note the bright spot (S) appearing on the cell surface close to the edge in the middle of 
the second image. It appears in less than 7 minutes and disappears soon after.
A similar one can be observed on the upper part of the last image. These features 
seem to be linked to the ends of curved filaments. In many cases micrometer sized 
unidentified nodes were found on stress fibres. They might be large protein complexes 
attached to F-actin. 

FIG.\ \ref{fig2} displays the height-time curves captured on the leading edge (c) 
and close to that on the cell body (d). Curves show increased motility especially on 
the long (several seconds) time scale. As a consequence, the saturation effect of the 
height-height correlation function disappears on this scale (FIG.\ \ref{fig3}). 
The speed of fast fluctuations is higher, as well. (See Table \ref{table1}.) We 
suppose that the observed increase in vertical fluctuations is an outcome of actin 
polymerization-depolymerization processes at the leading edge. Surprisingly, the 
height-height correlation function shows a higher level of fluctuations farther 
from the leading edge. It can be a result of actin depolymerization 
processes well behind the edge or an increased temporal motility of this region which 
has to follow the edge.

\section{Discussion}

The analysis of height fluctuations acquired at different locations 
allows a sensitive monitoring of the motility of cellular components. Both
actin-myosin based contractions and actin polymerization-based
filopodial and lamellipodial protrusions can be examined by this method. 
We found a correlation between the characteristics of vertical fluctuation 
and organized lateral locomotion. 

We explain the observed 5 Hz pulsation of a cell with the periodic contractions 
of stress fibres. This type of oscillation cannot be easily identified by other 
techniques due to its low amplitude. Although the frequency of mechanical pulsation of cardiomyocytes is in the 
same frequency range ($\sim$1.25 Hz), its amplitude is 2 
orders of magnitude higher \cite{a16}. Spontaneous oscillatory contractions of 
muscle fibers with a period of a few seconds are widely known for several
years (e.g. \cite{a20}). Theoretical models can explain spontaneous 
oscillation under certain conditions \cite{a21}).  

Slow pulsation of non-muscle cells has been observed in several cases. 
Microtubule depolymerization can induce rhythmic actomyosin-based contractility with a period of 
$\sim$50 s in fibroblasts \cite{a22} and oscillatory activity in the cortical
microfilament system of lymphoblasts \cite{a23}. Shape oscillations of leukocytes
driven by cyclic actin polymerization has been studied by several groups \cite{a24}. 
The period of this process is about  $\sim$8 s. 

The cortical tension of non-muscle cells generated by myosin-II can drive a 
change of shape \cite{a25}. Myosin molecules cycle about 5 times in a second 
in muscle \cite{a26}. Based on the above mentioned facts, we think that a synchronized 
behavior of myosin molecules in stress fibres may cause the observed pulsation. 
Myosin synchronization has been theoretically predicted close to the 
isometric condition in highly organized actin structures \cite{a27,a28}. Further 
experiments are needed to elucidate the background of this phenomenon. Using 
drugs affecting a specific system of the cytoskeleton will help to distinguish 
their roles in the nanometer scale fluctuations of cells. 

\section{Acknowledgments}

We thank El\H{o}d M\'ehes, Andr\'as Czir\'ok, Bal\'azs Heged\H{u}s for discussions 
and help in cell culturing, Jeremy Ramsden for consulting on the presentation 
of our results and Imre Der\'enyi for discussions in the field of myosin motors.

\widetext
\begin{table*}
\begin{tabular}{rlccccc}
\multicolumn{2}{c}{Curve} & Starting Slope [nm/s] & r & Saturation value $\pm$ SD [nm] & n & Lateral Velocity $\pm$ SD [nm/s]
 \\
\hline
Quiescent cell    & a & 6.3   & 0.94  & 10.5 $\pm$ 0.2 & 10 & 2.3 $\pm$ 0.4 \\
                  & a' & 8.4   & 0.97  & 17.0 $\pm$ 0.2 & 10 & 3.2 $\pm$ 0.5
 \\               & b & 16.7  & 0.97  & 28.1 $\pm$ 0.3 & 8  & 11.4 $\pm$ 2.5
 \\
\hline
Cell in motion    & a & 20.3  & 0.998 & --             & 10 & 5.3 $\pm$ 2.9
 \\
                  & b & 28.7  & 0.999 & --             & 9  & 5.3 $\pm$ 2.9
\end{tabular}
\caption{Comparison of the lateral velocity and parameters characterizing vertical fluctuations of different locations on the quiescent and the motile cell. Parameters of the height-height correlation function are calculated on the basis of curves presented in FIG.\ \ref{fig3}. Starting slope and saturation value were determined by linear fitting in the (0.3 s, 1 s) and (3 s, 5 s) intervals respectively, r: correlation coefficient of fitting, SD: Standard Deviation, n: number of measured height-time curves.
}
\label{table1}
\end{table*}

\begin{figure}
\caption{Shaded deflection mode images with a 4.5 minutes difference showing stable actin-myosin cables at the rear of a quiescent cell. Arrow in the lower left corner indicates the direction of motion. SF: stress fibres.
}
\label{fig1}
\end{figure}

 \begin{figure}
\caption{Typical vertical fluctuations measured on quiescent and motile cells. Graph (a) and (b) belong to the quiescent, (c) and (d) to the motile cell. We recorded (a) on the rear edge displayed in FIG.\ \ref{fig1}. (b) was measured on a typical retracting triangular shaped contact (image not shown) of the cell at the rear edge. This contact to the support was broken a few minutes after recording the vertical fluctuations. (c) and (d) were registered on the leading edge shown in FIG.\ \ref{fig5} and close to that on the cell body respectively.
}
\label{fig2}
\end{figure}

\begin{figure}
\caption{Averaged height-height correlation functions of fluctuations measured on the surface of the quiescent and the motile cell. Curves (a, n=10); (b, n=8); (c, n=10) and (d, n=9) are the corresponding correlation functions of vertical fluctuations shown in FIG.\ \ref{fig2}. Curve (a', n=10) was registered on the middle region of the quiescent cell. There is an apparent difference between the behavior of curves belonging to the quiescent and the motile cell. Saturation disappears on the scale of several seconds in case of the motile cell. This fact indicates the presence of vertical motility on this time scale. Significant difference between graph (a) and (b) is attributed to the dynamics of the retracting contact at the rear edge. See the value of lateral velocity of locations at each curve. 50 Hz noise on curves can be observed.
}
\label{fig3}
\end{figure}

\begin{figure}
\caption{Power spectrum (n=10) of height fluctuations measured at the location shown in FIG.\ \ref{fig1}. The peak at 4.9 Hz can be found in each power spectrum of height fluctuations captured on the surface of the quiescent cell. The power spectra of height fluctuations of cells without apparent stress fibres nearby lack this peak.
}
\label{fig4}
\end{figure}

\begin{figure}
\caption{Consecutive shaded deflection mode images of a leading edge. Approximately 7 minutes elapsed between images. Arrow indicates the direction of motion. Note the bright spot (S) appearing on the cell surface close to the edge in the middle of the second image. L: lamellipodium, F: filopodium. Contour lines (extreme right) display the forward motion of the edge. The standard deviation of lateral velocity was found to be higher than in the case of a less mobile edge due to extensions growing with high speed, such as the extension on the lower region of the last contour line corresponding to the right hand image.
}
\label{fig5}
\end{figure}


\begin{thebibliography}{99}

\bibitem{a1} L. M. Machesky and A. Hall, J.\ Cell Biology {\bf 138,} 913-926 (1997)

\bibitem{a2} H. Chen, B. W. Bernstein and J. R. Bamburg, TIBS {\bf 25,} 19-23 (2000)

\bibitem{a3} G. G. Borisy, T. M. Svitkina, Curr.\ Opin.\ Cell.\ Biol.\ {\bf 12} (1), 104-12 (2000) 

\bibitem{a4} L. Mahadevan and P. Matsudaira, Science {\bf 288,} 95-99 (2000) 

\bibitem{a5} Ch. S. Peskin, G. M. Odell and G. F. Oster, Biophys.\ J.\ {\bf 65,} 316-24 (1993) 

\bibitem{a6} A. Czir\'ok, K. Schlett, E. Madar\'asz, and T. Vicsek, Phys.\ Rev.\ Lett.\ {\bf 81} (14), 3038-41 (1998)

\bibitem{a7} B. Heged\H{u}s, A. Czir\'ok, I. Fazekas, T. B\'abel, E Madar\'asz, and T. Vicsek, J.\ Neurosurgery {\bf 92} (3) 428-34 (2000)

\bibitem{a8} Zs. K\"ornyei, A. Czir\'ok, T. Vicsek, E. Madar\'asz, J.\ Neurosci.\ Res.\ {\bf 61,} 421-29 (2000)

\bibitem{a9} H. J. Butt, E. K. Wolff, S. A. Gould, B. Dixon Northern, C. M. Peterson, P. K. Hansma, J.\ Struct.\ Biol.\ {\bf 105,} 54-61 (1990)

\bibitem{a10} E. Henderson, P. G. Haydon, D. S. Sakaguchi, Science {\bf 257,} 1944-46 (1992)

\bibitem{a11} V. Parpura, P. G. Haydon, E. Henderson, J.\ Cell Sci. {\bf 104,} 427-32 (1993)

\bibitem{a12} Ch. Rotsch, K. Jacobson and M. Radmacher, PNAS {\bf 96,} 921-26 (1999) 

\bibitem{a13} C. Rotsch, M. Radmacher, Biophys.\ J.\ {\bf 78} (1), 520-35 (2000)

\bibitem{a14} U. G. Hofmann, C. Rotsch, W. J. Parak, M. Radmacher, J.\ Struct.\ Biol.\ {\bf 119} (2), 84-91 (1997)

\bibitem{a15} M. Lekka, P. Laidler, D. Gil, J. Lekki, Z. Stachura, A. Z. Hrynkiewicz, Eur.\ Biophys.\ J.\ {\bf 28,} 312-16 (1999)

\bibitem{a16} J. Domke, W. J. Parak, M. George, H. E. Gaub, M. Radmacher, Eur.\ Biophys.\ J.\ {\bf 28,} 179-86 (1999) 

\bibitem{a17} S. W. Schneider, K. C. Sritharan, J. P. Geibel, H. Oberleithner and B. P. Jena, PNAS {\bf 94,} 316-21 (1997) 

\bibitem{a18} H. Haga, M. Nagayama, K. Kawabata, E. Ito, T. Ushiki, T. Sambongi, J.\ Electron Microsc.\ (Tokyo) {\bf 49} (3), 473-81 (2000) 

\bibitem{a19} C. A. Sh\"onenberger, J. H. Hoh, Biophys.\ J.\ {\bf 67} (2), 929-36 (1994)

\bibitem{a20} K. Yasuda, Y. Shindo, S. Ishiwata, Biophys.\ J.\ {\bf 70} (4), 1823-9 (1996)   

\bibitem{a21} F. J\"ulicher, A. Ajdari, J. Prost, Rev.\ Mod.\ Phys.\ {\bf 69} (4), 1269-81 (1997)

\bibitem{a22} O. J. Pletjushkina, Z. Rajfur, P. Pomorski, T. N. Oliver, J. M. Vasiliev, K. A. Jacobson, Cell.\ Motil.\ Cytoskeleton {\bf 48} (4), 235-44 (2001)

\bibitem{a23} M. Bornens, M. Paintrand, C. Celati, J. \ Cell. \ Biol.\ {\bf 109} (3), 1071-83 (1989)

\bibitem{a24} M. U. Ehrengruber, D. A. Deranleau, T. D. Coates, J.\ Exp.\ Biol.\ {\bf 199,} 741-47 (1996) 

\bibitem{a25} C. Pasternak, J. A. Spudich, E. L. Elson, Nature {\bf 341,} 549-51 (1989)

\bibitem{a26} B. Alberts, D. Bray, J. Lewis, M. Raff, K. Roberts, J. D. Watson, Molecular Biology of the Cell, 3rd Edition, p. 853, Garland Publishing, New York (1994)

\bibitem{a27} T. A. J. Duke, PNAS {\bf 96,} 2770-75 (1999)

\bibitem{a28} T. Duke, Philos. Trans. \ R. \ Soc. \ Lond. \ B Biol.\ Sci. \ {\bf 355} (1396), 529-38 (2000)
  
\end{thebibliography}
\end{document}